\documentclass[fleqn,10pt]{wlscirep}
\title{Universality and critical behavior of the dynamical Mott transition in a system with long-range interactions}

\author[1,*]{Louk Rademaker}
\author[2]{Valerii V. Vinokur}
\author[2,3]{Alexey Galda}
\affil[1]{Kavli Institute for Theoretical Physics, University of California Santa Barbara, CA 93106, USA}
\affil[2]{Materials Science Division, Argonne National Laboratory, Argonne, Illinois
	60439, USA}
\affil[3]{James Franck Institute, University of Chicago, Chicago, Illinois 60637, USA}

\affil[*]{louk.rademaker@gmail.com}

\begin{abstract}
We study numerically the voltage-induced breakdown of a Mott insulating phase in a system of charged classical particles with long-range interactions. At half-filling on a square lattice this system exhibits Mott localization in the form of a checkerboard pattern. We find universal scaling behavior of the current at the dynamic Mott insulator-metal transition and calculate scaling exponents corresponding to the transition. Our results are in agreement, up to a difference in universality class, with recent experimental evidence of dynamic Mott transition in a system of interacting superconducting vortices.\end{abstract}
\begin{document}

\flushbottom
\maketitle
%
\thispagestyle{empty}


\section*{Introduction}

Materials exhibiting electric field-driven (dynamic) Mott metal-insulator transition (MIT) have a high potential for replacing semiconductors due the unique property of controllable energy gap, making them extremely promising for future low-energy electronics. While the physical mechanism behind dynamic MITs in most experimental systems is still unclear\cite{Poccia2015,Li:2015hi}, several theories have been proposed, including avalanche breakdown\cite{Guiot:2013ih,Stoliar:2013dd} and Schwinger-Landau-Zener tunneling\cite{Oka:2012ht,Oka:2010ku,Oka:2003hr,Schwinger:1951is} associated with parity-time symmetry-breaking\cite{Tripathi2016}.

In this Letter we investigate a classical system of long-range interacting charged particles experiencing voltage-induced Mott MIT near half-filling on a square lattice at small temperatures. At sufficiently low applied voltage across the system, particles arrange themselves in a checkerboard pattern. Strong inter-particle interaction impedes any motion at exactly half-filling, forming a Mott insulator. This state can be broken by either increasing temperature (thermodynamic transition) or by applying sufficiently strong external electric field or voltage, causing a dielectric breakdown characterized by finite conductivity. We observe scaling behavior at the dynamic Mott MIT with differential conductivity of the system being a universal function of $|V - V_c|/|f - f_c|^\epsilon$, where $V$ is the applied voltage, $|f - f_c|$ is deviation from commensurate particle density, $f_c = 0.5$, $\epsilon$ is some scaling exponent, and $V_c$ is the critical amplitude of applied voltage inducing the transition.

\section*{Model and simulation details}
We study a lattice gas model with long-range Coulomb interactions on a two-dimensional square lattice, with energy of the system given by the expression
\begin{equation}
	E_C = \sum_{ij}\frac{1}{r_{ij}} (n_i - \bar n)(n_j - \bar n)\,,
\end{equation}
where $n_i = 0, 1$ represents the particle occupation number of site $i$, and $r_{ij}$ is the distance between sites $i$ and $j$. At low temperatures $T \rightarrow 0$ and half-filling $f = 0.5$ the system displays Mott localization and displays a corresponding checkerboard charge order pattern.\cite{Rademaker:2013jg} To realize the dynamic Mott transition we apply an electric field along the $x$-direction,
\begin{equation}
	E_V = - \sum_i V x_i n_i.
	\label{Efield}
\end{equation}
Here $V$ is the electric potential ans $x_i$ is the $x$-coordinate of the site $x$. In the remainder of this section we will describe how we simulated this model.

\begin{figure*}[!thb]
	\centering
	\includegraphics[width=0.5\textwidth]{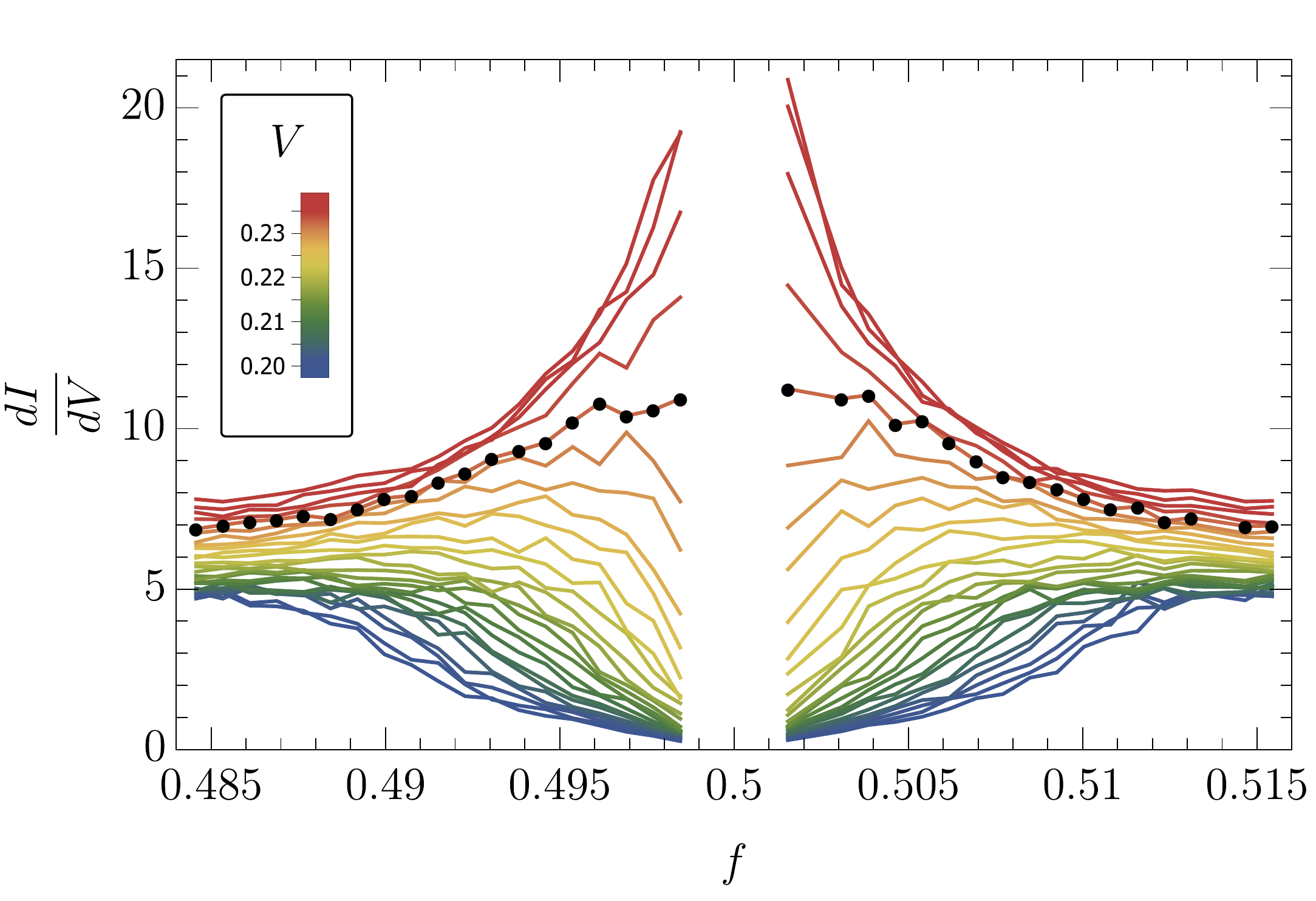}
	\caption{$dI/dV$ curves in the vicinity of half-filling. The critical voltage, $V_c = 0.233 \pm 0.005$, is determined from the criterion $\frac{d}{df}\!\left.\left( \frac{dI}{dV}\right)\right|_{V = V_c} = 0$. The corresponding $dI/dV$ curves are marked by black solid dots.}
	\label{Fig1}
\end{figure*}

We consider a square lattice of linear dimension $L$ with periodic boundary conditions and take into account the long-range nature of Coulomb interaction by employing the Ewald summation method\cite{Toukmaji96}. In the two-dimensional case, the Ewald sum is split into a constant energy term and two rapidly converging sums over real and reciprocal space, correspondingly:
\begin{align}
	E_{ij} &= -\frac{2\alpha}{\sqrt{\pi}}\delta_{ij} + \sum_{\mathbf{n}}\frac{\text{Erfc}\left( \left| \mathbf{r_{ij}} + \mathbf{n}\,L\right| \alpha\right)}{\left| \mathbf{r_{ij}} + \mathbf{n}\,L\right|}
	+ \sum_{\mathbf{m} \neq 0}\frac{(2\pi)^2}{\pi L^2}\frac{e^{-\pi^2|\mathbf{m}|^2/(2\pi\alpha)^2}}{|\mathbf{m}|^2}\cos(\mathbf{m}\cdot\mathbf{r_{ij}})\,,
\end{align}
where $\mathbf{n}$ and $\mathbf{m}$ are integer vectors.

We performed Monte Carlo simulation with the heat-bath local update algorithm. At each computational step one randomly chosen particle is proposed to move to one of its neighboring sites. The acceptance probability is $P_{i \to j} = \frac{e^{-\Delta E/T}}{1 + e^{-\Delta E/T}}$, where $\Delta E$ is the corresponding change in energy. To promote particle conductivity, the electric field applied along the $x$ axis of equation (\ref{Efield}) is modeled by including in $\Delta E$ a lowering (raise) by $V$ if the suggested move is to the right (left). Note that due to periodic boundary conditions, the total energy of a particle configuration is only defined up to a multiple of $VL$. Because the Monte Carlo simulation is only dependent on energy \textit{differences}, however, this does not pose a problem.

In what follows we will assume that all particles have a unit charge, and take the lattice spacing as a unit of distance, making $V$ also a measure of applied electric field in dimensionless units. To calculate the current generated during simulations, we count the number of particles crossing the $x = 0$ line per single Monte Carlo sweep, where one sweep is defined as $L^2$ proposed moves. Note that the current measured this way is limited by the number of particles present in the system and, therefore, has an unphysical upper bound. This fact limits the validity of our approach to studying particle conductivity at low voltages. To what extent such saturation effects influence the results can be probed by checking the acceptance rate of moves in the $x$ direction.

Each complete Monte Carlo simulation has been performed at a fixed overall particle density $f$ in two stages. First, the system's thermal equilibrium state was reached by annealing in the absence of an external applied voltage at temperature $T = 0.04$, followed by incremental increases in the applied electric field by ${dV = 1/600}$ and measurements of particle conductivity at each voltage.

Our results were obtained for the lattice sizes ${L = 36}$ with $628$ to $668$ particles, corresponding to the range of densities ${f \approx [0.485, \,0.515]}$. Each data point presents an average over $2.88$ million Monte Carlo sweeps. We also studied the system at smaller sizes $L=12, 24$ which gave similar results, however, for clarity we will only present the data for the largest lattice size $L=36$. Differential conductivity data, $dI/dV$, were obtained from $IV$ curves by a five-point stencil.

\begin{figure}[!thb]
	\centering	
	\includegraphics[width=0.48\textwidth]{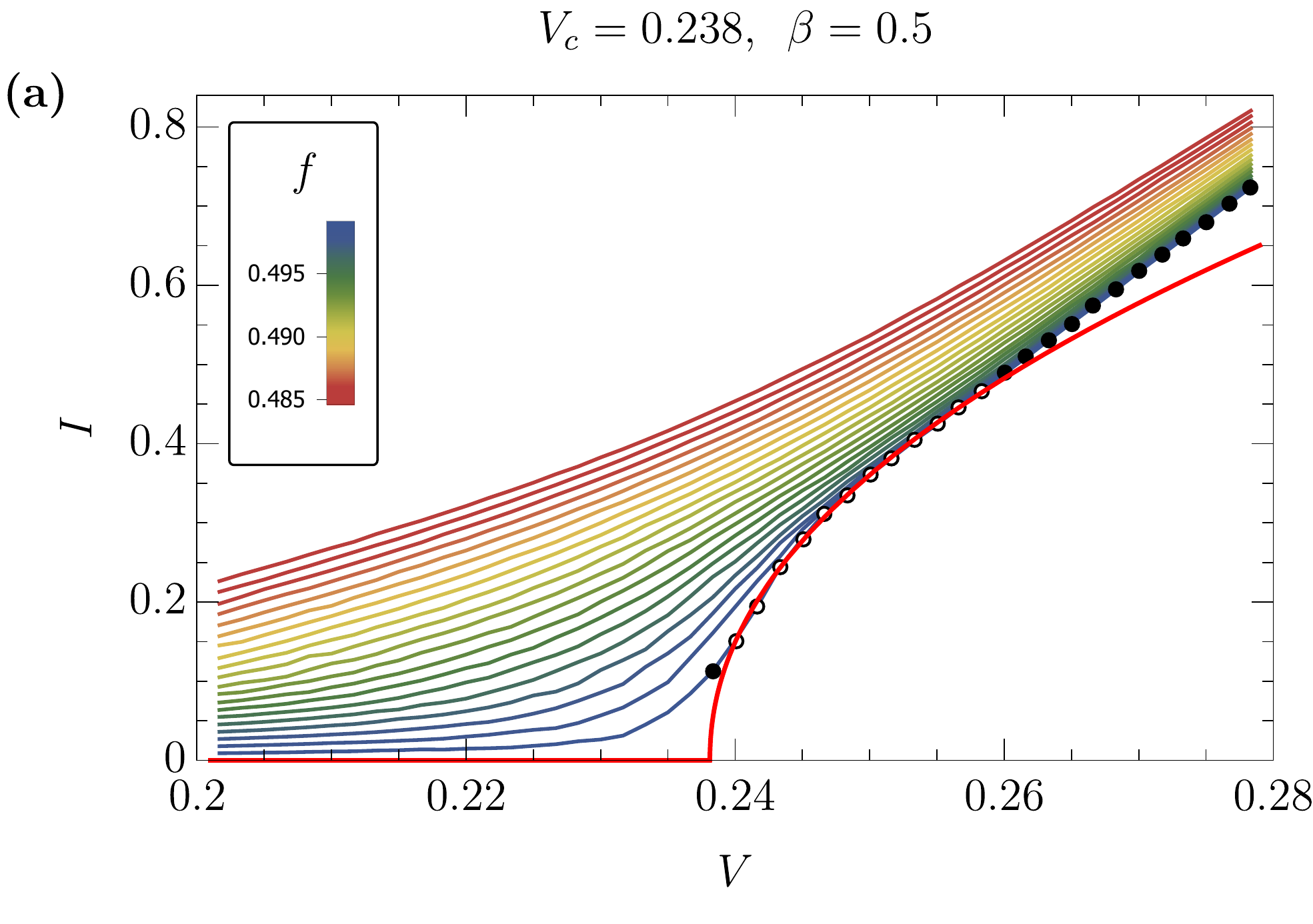}
	\includegraphics[width=0.48\textwidth]{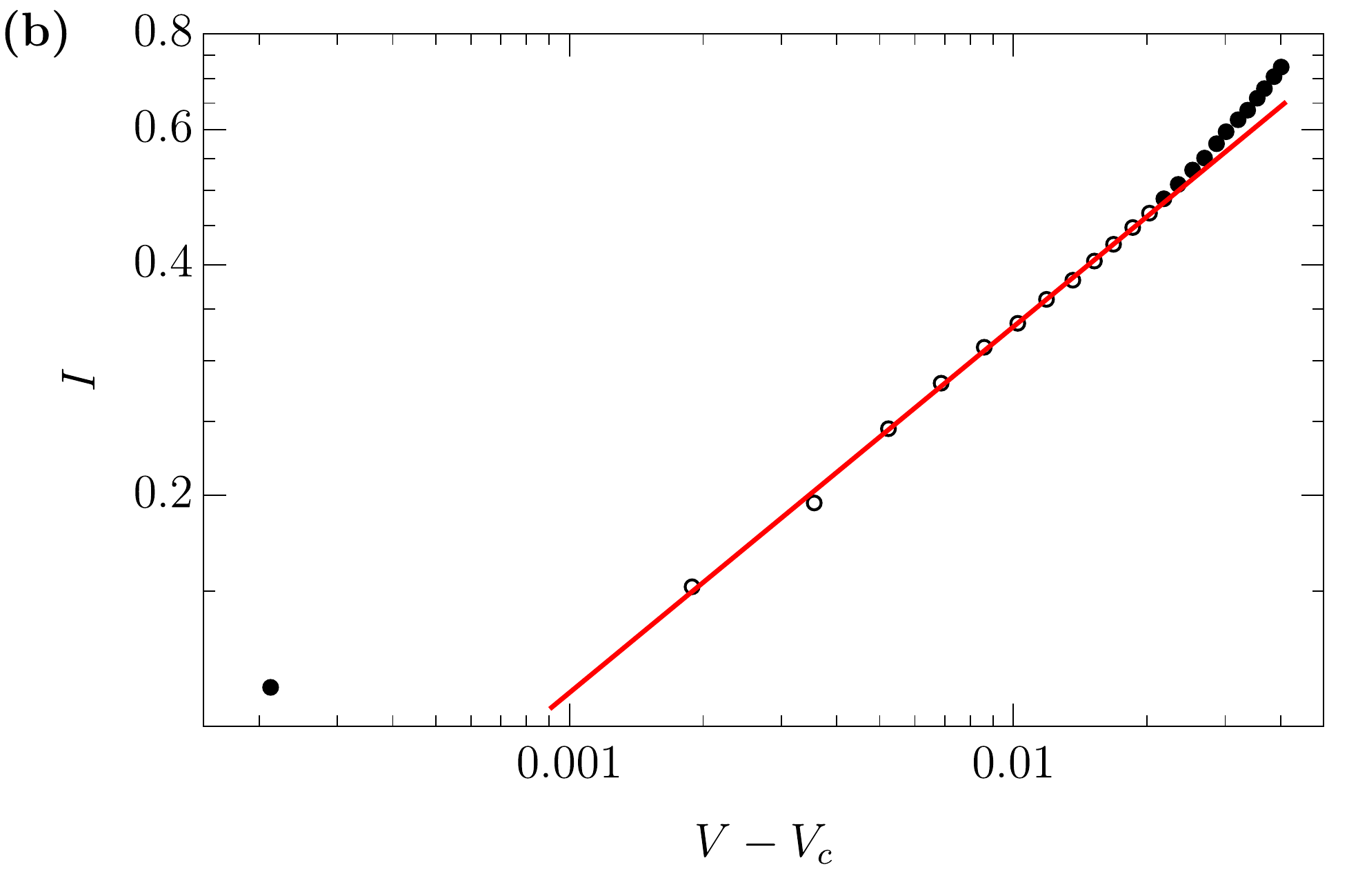}
	\caption{(a) Current, $I$, as a function of applied voltage, $V$, near the dynamic Mott MIT for a range of densities near $f_c$, $f_c - f < 0.015$. Red line represents the best fit of data in the form~(\ref{scaling1}) for $f = 0.499$. Solid (empty) circles represent data points outside (inside) of the critical regime. (b) Current, $I$, as a function of deviation of applied voltage from the critical value, $V - V_c$, for $V_c = 0.238$. Red straight line fits data points in the critical regime with power-law scaling above the transition (for $V > V_c$).}
	\label{Fig2}
\end{figure}

\begin{figure}[!hb]
	\centering
	\includegraphics[width=0.48\textwidth]{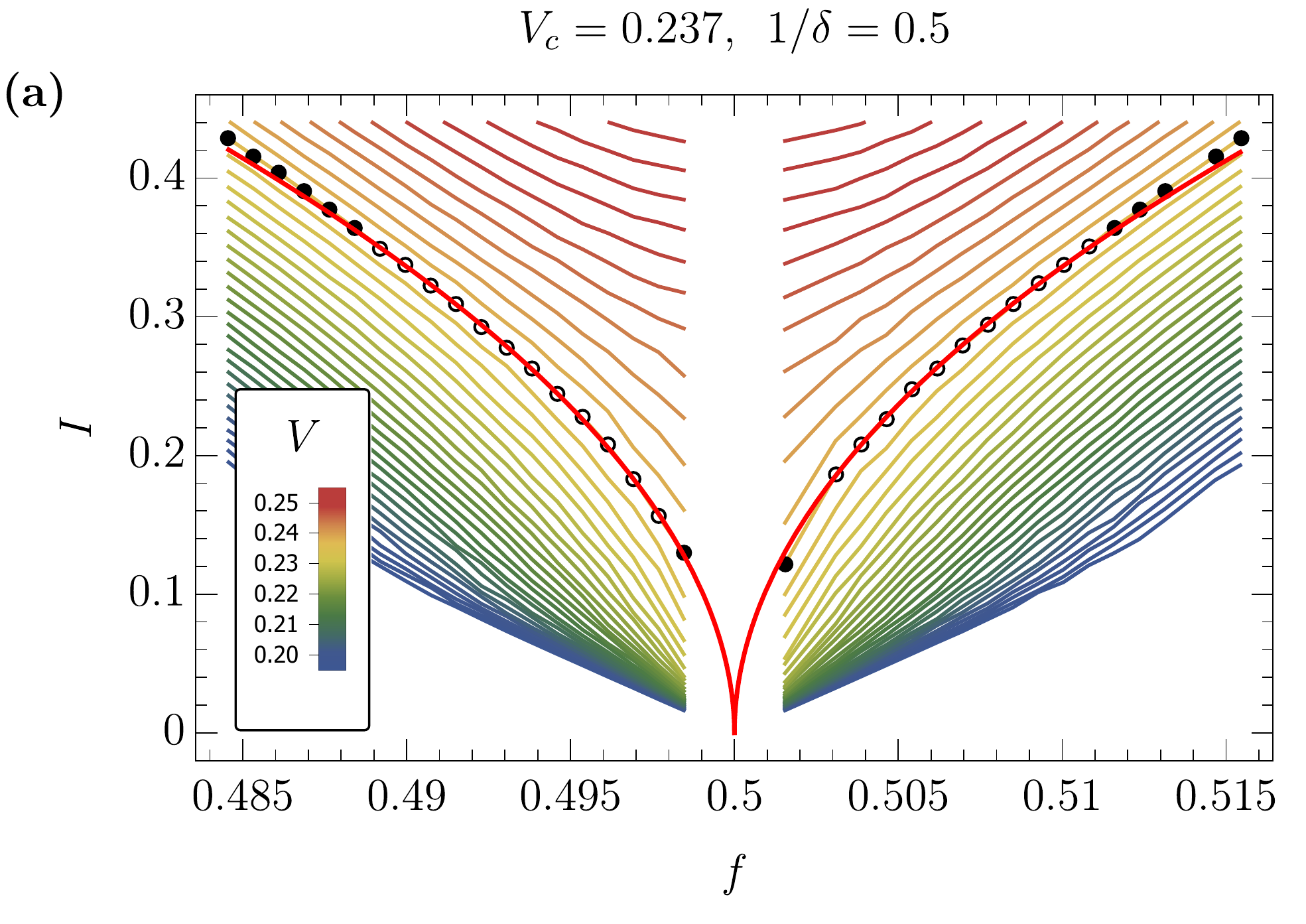}
	\includegraphics[width=0.48\textwidth]{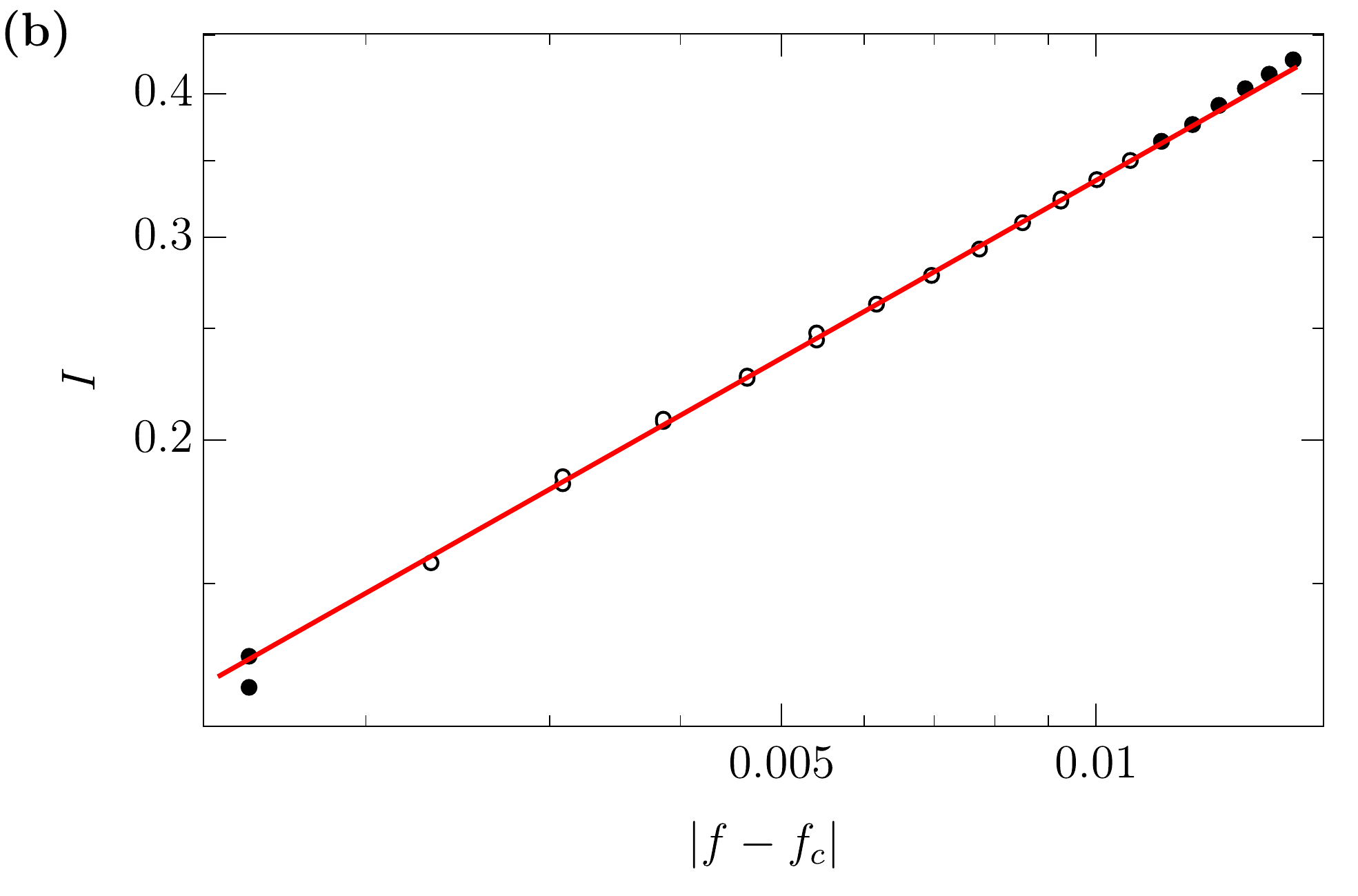}
	\caption{(a) Current, $I$, as a function of particle particle density, $f$, for a range of applied voltages, $V$. Red lines show the best fit of $I(V_c, f)$ curves, Eq.~(\ref{scaling2}) in the critical regime near $f_c$. (b) Current as a function of deviation from critical particle density, $I(|f - f_c|)$ at $V = 0.237$. Solid and empty circles represent data points outside and inside of the critical regime with power-law scaling~(\ref{scaling2}), correspondingly.}\label{Fig3}
\end{figure}

\begin{figure}[!hb]
	\centering
	\includegraphics[width=0.48\textwidth]{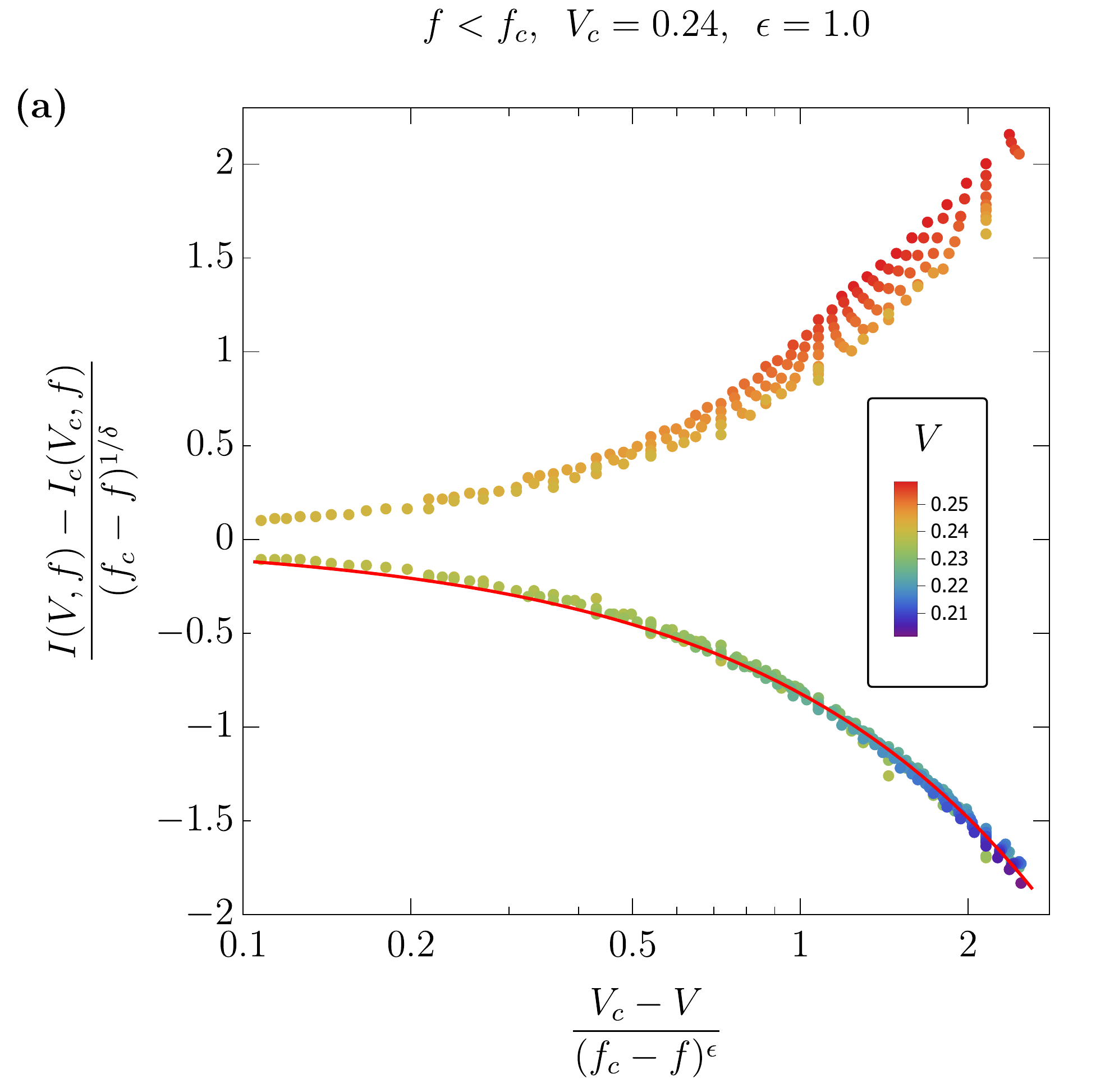}
	\includegraphics[width=0.48\textwidth]{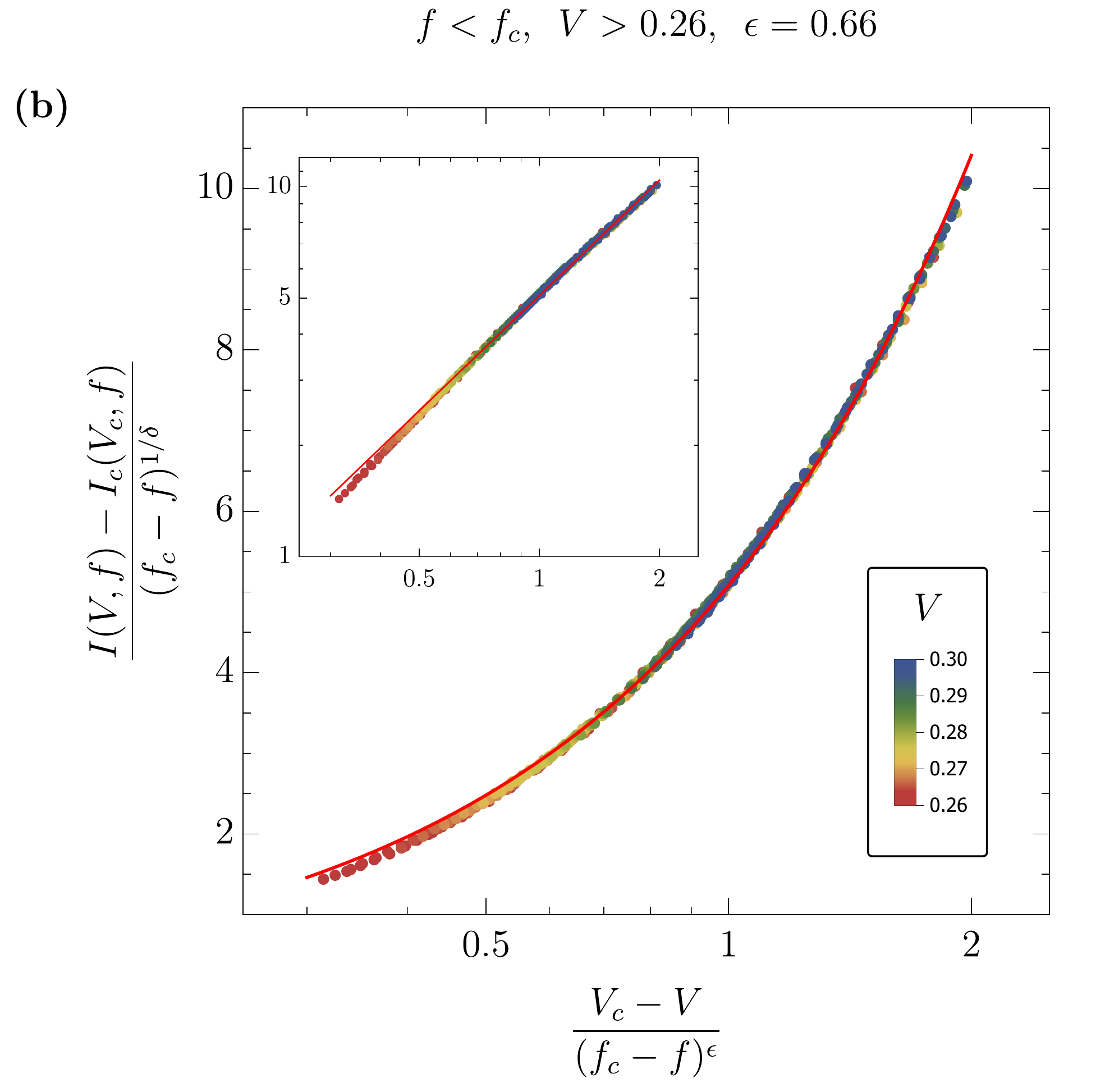}
	\caption{(a) Universal scaling of current according to Eq.~(\ref{scaling}). Lower branch corresponds to applied voltages below critical, $V < V_c$, while the upper branch shows a small range of data at voltage directly above critical, $0.24 < V < 0.25$. Fitting analysis results in the scaling exponent $\epsilon = 1.0$ neat Mott MIT. (b) Universal scaling of current at $V > 0.26$, where finite number of particles in the system causes a different from the Mott MIT scaling behavior, with a scaling exponent $\epsilon = 0.66$.}\label{Fig4}
\end{figure}

\begin{figure}
	\centering
	\includegraphics[width=0.5\textwidth]{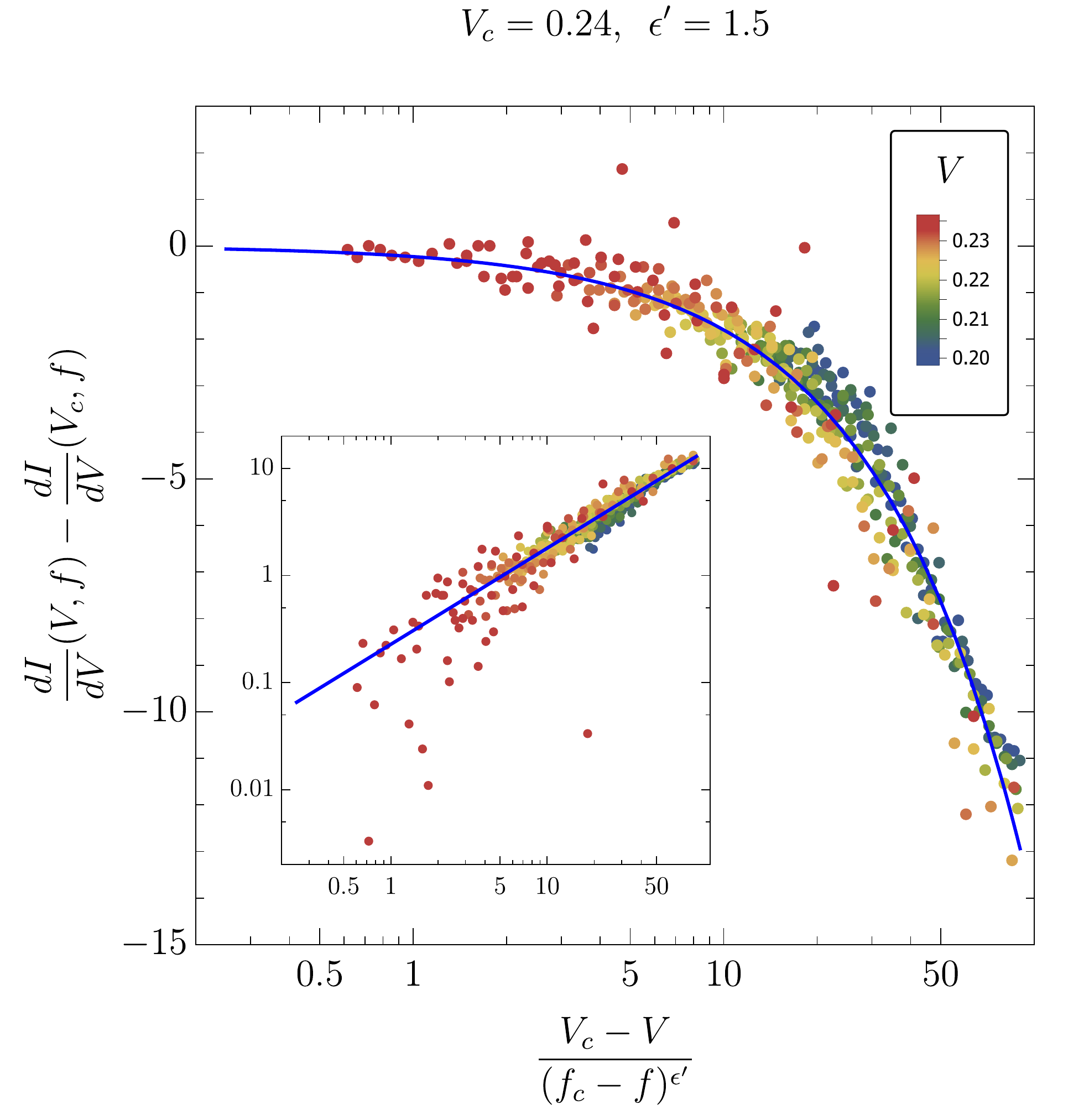}
	\caption{Scaling of the differential conductivity $dI/dV$ around the dynamical critical point at $V_c = 0.238$ and $f_c = 0.5$. Following Eq.~(\ref{scaling}), we find $\epsilon = 1.34$ and $\mu = 1.0$.}\label{Fig5}
\end{figure}

\section*{Results}

From the calculated $dI/dV$ curves, we find the critical voltage $V_c = 0.233 \pm 0.005$ near half-filling, see Fig.~\ref{Fig1}, below which the system behaves as an insulator, and above which it is conducting with significant non-zero current flowing through the system. Simulations revealed that in the immediate vicinity of $f = 0.5$ (region of absent data in Fig.~\ref{Fig1}), particle current is mostly generated by an avalanche-like motion of melted clusters in the checkerboard arrangement and not by excitation of individual particles driving the dynamic MIT. We found that current, as a function of particle density $f$, experiences discontinuity at $f = 0.5$, i.e. $\lim_{f \to 0.5} I(f) \neq I(0.5)$. A study of this collective effect lies outside the scope of the present Letter and will be considered elsewhere.

To show that the field-driven MIT considered here is a phase transition, we study the behavior of charge current generated in the system by applied transverse voltage. As expected for a phase transition, we observe power-law scaling of the current both as a function of applied voltage,
\begin{equation}\label{scaling1}
		I(V, f_c) = \begin{cases}
		(V - V_c)^\beta, & V \geq V_c \\
		\,0, & V < V_c
		\end{cases}\,,
\end{equation}
and as a function of particle density,
\begin{equation}\label{scaling2}
	I(V_c, f) \sim |f - f_c|^{1/\delta}\,,
\end{equation}
see Figs.~\ref{Fig2} and ~\ref{Fig3}, correspondingly.

In Fig.~\ref{Fig2}a we plot the $IV$ curves for ${f < 0.5}$,
revealing a sharp increase in measured current above the critical voltage $V_c$, which becomes progressively more pronounces as particle density $f$ approaches $f_c$. Nonlinear regression analysis was performed for the $I(f_c, V)$ data\cite{Markovic1999} to achieve the best fit to the function~(\ref{scaling1}) and resulted in the scaling exponent $\beta = 0.5 \pm 0.1$ and critical voltage $V_c = 0.238 \pm 0.002$, which is in full agreement with $V_c$ determined based on the $dI/dV$ curves from Fig.~(\ref{Fig1}). The critical region corresponding to the power-law fitting~(\ref{scaling1}) was determined based on the extent of the linear range of $I(f_c, V - V_c)$ in double-logarithmic coordinates, see Fig.~\ref{Fig2}b. The nearest to $V_c$ data point seems to be largely affected by fluctuations of the measured current near the transition, and, together with the data at $V - V_c \gtrsim 0.02$ does not belong to the critical regime of the Mott MIT transition. In fact, we observe a noticeable deviation of the scaling exponent $\beta$ from the $0.5$ value at $V \geq 0.26$, which can be attributed to the onset of finite system size effects.

In the regime of fixed voltage, at $V = 0.237$ near the critical voltage value, we find the power-law scaling of $I(f)$ in the form of equation~(\ref{scaling2}) in the vicinity of $f_c = 0.5$, see Fig.~\ref{Fig2}a, with the critical exponent $1/\delta = 0.5 \pm 0.1$. Fig.~\ref{Fig2}b reveals the extent of the critical regime with power-law scaling. The nearest to $f_c = 0.5$ data points appear to be affected by the avalanche physics, while at $|f - f_c| \gtrsim 0.01$ the power-law scaling starts to deviate from the investigated critical one, given by Eq.~(\ref{scaling2}).

Our main result comes from analysis of the whole set of data around the critical point, $\{V_c, f_c\}$, where we observe universal scaling behavior of the measured current in the following form:
\begin{equation}\label{scaling}
	I(V, f) - I_c(V_c, f) = |f - f_c|^{1/\delta} F_{\pm}\left( \frac{|V - V_c|}{|f - f_c|^{1/(\delta\beta)}}\right).
\end{equation}
$F_\pm (x)$ are the scaling functions in the metallic (${V > V_c}$) and insulating (${V < V_c}$) phases, correspondingly. It follows from the scaling relations~(\ref{scaling1})--(\ref{scaling2}) that $F_\pm (x \ll 1) \sim \text{const}$, and $F_+(x \gg 1) \sim x^\beta$.

We have performed scaling analysis of the $I(V, f)$ data in the form:

\begin{equation}\label{epsilon}
	\frac{I(V, f) - I(V_c, f)}{|f - f_c|^{1/\delta}} = F_{\pm}\left( \frac{|V - V_c|}{|f - f_c|^\epsilon}\right).
\end{equation}
The scaling parameters were obtained by maximizing the non-adjusted coefficient of determination, $R^2$, 
for the non-linear fit model~(\ref{epsilon}), which resulted in $1/\delta = 0.5$, $V_c = 0.24$ and $\epsilon = 1.0$, see Fig.~\ref{Fig4}a. The above values are in full agreement with the separate analysis based on Eqs.~(\ref{scaling1})--(\ref{scaling2}). We found that the scaling relation $\epsilon = (\delta\beta)^{-1}$, cf. Eqs.~(\ref{scaling}) -- (\ref{epsilon}), holds with remarkable accuracy.

Due to the current-saturation effect at $V \gtrsim 0.26$ caused by finite system size, 
scaling of the upper branch ($V > V_c$) in Fig.~\ref{Fig4}a is considerably less accurate. In fact, for $V > 0.26$ one obtains universal scaling behavior with $\epsilon = 0.66$, see Fig.~\ref{Fig4}b, which explains poor scaling in the transient voltage range between 0.24 and 0.26.

Following Ref. [\citen{Poccia2015}], we have analyzed differential conductance data, $dI/dV(V, f)$. We have also observed universal scaling behavior in the form
\begin{equation}\label{scalingD}
	\frac{dI}{dV}(V, f) - \frac{dI}{dV}(V_c, f) = G_{\pm}\left( \frac{|V - V_c|}{|f - f_c|^{\epsilon{^\prime}}}\right).
\end{equation}
The critical parameters determined based on the best fit to Eq.~(\ref{scalingD}) were found to be $V_c = 0.238$ and $\epsilon^\prime = 1.5$, see Fig.~\ref{Fig5}. It follows from Eq.~(\ref{scaling}) that the scaling exponent $\epsilon^\prime$ must satisfy the following relation:
\begin{equation}
	\epsilon^\prime = \frac{2 - \beta}{\delta\beta} = (2 - \beta)\epsilon\,,
\end{equation}
which for $\epsilon = 1$, $1/\delta = 0.5$ and $\beta = 0.5$ yields $\epsilon^\prime = 1.5$.
Note that only scaling of differential conductance on the lower branch ($V < V_c$) was performed due to the saturation effect at high voltages mentioned earlier, which limited the amount of scaling-suitable data points with reasonably small errors at $V > V_c$. Additionally, noise levels in $dI/dV$ data are significantly higher than in the raw $I(V, f)$ data due to numerically calculated first derivative.

\section*{Conclusions}

In a simple model of interacting classical particles with long-range interactions we have observed universality of critical behavior near the transition between the insulating Mott state with checkerboard order and the conducting liquid-like state above a critical value of applied voltage $V_c$. The main critical exponents, $\beta = 0.5, 1/\delta = 0.5$ and $\epsilon = 1.5$ satisfy to a remarkable accuracy scaling relations corresponding to a scaling form of the current and differential conductance around the critical point.

Our exponents differ from the ones in a dynamical MIT in a Josephson junction array\cite{Poccia2015}, where $\epsilon = 2/3$ at density $f=1$ and $\epsilon=1/2$ at $f=0.5$. We attribute the difference in universality classes of these two transitions to the form of interaction potential. While classical particles considered in the present Letter interact via $1/r$ Coulomb interaction, the interaction potential between vortices in Josephson junction arrays is logarithmic. An experimental realization of true Coulomb interactions is possible by taking an two-dimensional electron gas (2DEG) and apply an external periodic potential to mimic the square lattice.

Finally, it is worth mentioning that so far most of the work on a dynamic Mott transition has focused on disordered systems\cite{Middleton:1993,Ladieu:1996bo,Altshuler:2009ey,Ovadia:2009hi}. The transition there is fundamentally different as the main mechanism is depinning, rather than the reduction of the Mott gap. The dynamical Mott transition in clean systems, as first observed in Ref. [\citen{Poccia2015}], demands a much firmer theoretical understanding.




\section*{Acknowledgements}

We thank Ivar Martin and Lusine Khachatryan for fruitful discussions. L.\,R. is supported by the Dutch Science Foundation (NWO) through a Rubicon grant. V.\,V. and A.\,G. are supported by the U.S. Department of Energy, Office of Science, Materials Sciences and Engineering Division.

\section*{Author contributions statement}

L.R. did the Monte Carlo simulations, A.G. instigated the research and performed the data analysis. All authors contributed equally to the interpretation of the results and reviewed the manuscript.

\section*{Additional information}
The author(s) declare no competing financial interests.

\end{document}